\documentclass[12pt, final]{l4dc2023}

\title[Short Title]{Full Title of Article}
\usepackage{times}

\usepackage{bm}
\usepackage[capitalise]{cleveref}
\crefname{lemma}{Lemma}{Lemmas}
\usepackage{wrapfig}

\usepackage{xcolor}
\usepackage{fp, tikz, pgfplots}
\usetikzlibrary{arrows,shapes,backgrounds,patterns,fadings,matrix,arrows,calc,
	intersections,decorations.markings,
	positioning,arrows.meta}
\usepgfplotslibrary{fillbetween}
\usepgfplotslibrary{statistics}

\pgfplotsset{width=5\columnwidth /5, compat = 1.13,
	height = 60\columnwidth /100, grid= major,
	legend cell align = left, ticklabel style = {font=\scriptsize},
	every axis label/.append style={font=\small},
	legend style = {font={\tiny\sffamily}},title style={yshift=-7pt, font = \small} }

\newtheorem{assumption}{\bf{Assumption}}

\xdefinecolor{blueZ}{RGB}{0,77,163}

\DeclareMathOperator*{\argmax}{arg\,max}

\newif\ifarxiv
\arxivfalse

\title[Event-Triggered Learning with Computational Delays]{
Can Learning Deteriorate Control? Analyzing Computational Delays in Gaussian Process-Based Event-Triggered Online Learning
}

\author{%
	\Name{Xiaobing Dai}\thanks{Equal contribution.} \Email{xiaobing.dai@tum.com}\\
	\Name{Armin Lederer}\footnotemark[1] \Email{armin.lederer@tum.com}\\
	\Name{Zewen Yang} \Email{zewen.yang@tum.com}\\
	\Name{Sandra Hirche} \Email{hirche@tum.com}\\
	\addr Chair of Information-oriented Control, Technical University of Munich, D-80333 Munich, Germany%
}

\begin{document}

\setlength{\abovedisplayskip}{4.0pt}
\setlength{\belowdisplayskip}{4.0pt}
\setlength{\floatsep}{7pt}
\setlength{\textfloatsep}{7pt}

\maketitle

\vspace{-0.5cm}

\begin{abstract}%
	When the dynamics of systems are unknown, supervised machine learning techniques are commonly employed to infer models from data. Gaussian process (GP) regression is a particularly popular learning method for this purpose due to the existence of prediction error bounds. Moreover, GP models can be efficiently updated online, such that event-triggered online learning strategies can be pursued to ensure specified tracking accuracies. However, existing trigger conditions must be able to be evaluated at arbitrary times, which cannot be achieved in practice due to non-negligible computation times. Therefore, we first derive a delay-aware tracking error bound, which reveals an accuracy-delay trade-off. Based on this result, we propose a novel event trigger for GP-based online learning with computational delays, which we show to offer advantages over offline trained GP models for sufficiently small computation times. Finally, we demonstrate the effectiveness of the proposed event trigger for online learning in simulations.
\end{abstract}

\begin{keywords}%
  Gaussian process regression, learning-based control, computational delay, event-triggered learning, online learning
\end{keywords}

\section{Introduction}

The control of systems with unknown or uncertain dynamics is a challenging problem arising in many applications such as underwater vehicles \citep{yanTrajectoryTrackingControl2019}, electric machines \citep{yin2022implementation} and soft robotics \citep{gao2023quasi}.
In order to overcome this challenge, supervised machine learning methods are frequently applied to learn models of the unknown components. In particular when safety or performance guarantees are required for control, Gaussian process (GP) regression \citep{williams2006gaussian} is popular for model inference.

GP models have been employed together with a wide range of control techniques such as feedback linearization \citep{Greeff2021}, sliding mode control \citep{lima2020sliding}, control barrier function approaches \citep{Jagtap2020}, and model predictive control \citep{Maiworm2021}. 
By combining Lyapunov stability theory \citep{khalil2015nonlinear} with GP prediction error bounds \citep{srinivas2012information, lederer2019uniform}, data-dependent safety and performance guarantees can often be derived. 
When no data is available before system operation, GP models of the dynamics can be efficiently learned online via iterative model updates \citep{Nguyen-Tuong2009}. 
While this can be done using time-triggered model updates \citep{Meier2016a}, event-triggered learning \citep{Solowjow2018} offers the beneficial property that it can ensure the satisfaction of prescribed accuracy guarantees, such that desired tracking and safety guarantees of the controller are ensured by design \citep{umlauft2019feedback,castaneda2022probabilistic}. 
However, these guarantees crucially rely on the capability to evaluate the triggering conditions at arbitrary times, which is generally not possible in practice, in particular since GP model updates and predictions take non-negligible computation time \citep{lederer2021gaussian}. 
Therefore, GP models introduce a computational delay in control loops.\looseness=-1

While the control of delayed systems is a well-understood problem in classical control theory \citep{fridman2014introduction}, it has received little attention in the context of learning-based control. When linear regression is employed with nonlinear features, different forms of delays such as state \citep{ho2005adaptive} and input delays \citep{li2018neural} have been analyzed. However, these results  do not extend beyond linear regression, such that they are not applicable to GP models, even though computational delays can be modeled as time-varying input delay. Delays in GP models have, to the best of our knowledge, only been considered in the context of sampled-data control, e.g., realized through a self-triggered update of control inputs \citep{dhiman2021control}. 
However, these approaches do not consider the problem of designing event-triggers for model updates. Moreover, they only take into account the effect of zero-order hold GP predictions on their safety guarantees, but ignore their delayed availability due to the required computation time. 
Therefore, modeling learning-based control as sampled-data system does not sufficiently address the challenges caused by computational delays.\looseness=-1

In this work, we consider the problem of designing an event trigger for GP-based online learning with computational delays. For this purpose, we quantitatively analyze the effect of computational delays on tracking error bounds, which reveals a trade-off between prediction accuracy and computational delay. Based on this result, a trigger condition for online model updates is designed, such that a specified tracking error is guaranteed. Since online model updates generally increase the computation time, we show that event-triggered learning offers advantages when no sufficiently accurate GP model can be trained offline. Finally, numerical simulations demonstrate the practical importance of the derived theories and the effectiveness of the proposed online learning method.\looseness=-1

The remainder of this paper is structured as follows. \cref{section_problem} defines the problem setting. The control performance with computational delays is analyzed in \cref{section_delayControl}. In \cref{section_ET}, the event-triggered model update strategy is proposed. Numerical simulations are provided to illustrate the proposed method in \cref{section_simulation}.
Finally, \cref{section_conclusion} concludes this paper.

\section{Problem Setting}\label{section_problem}

In this paper, we consider plants described by $m$-th order dynamical systems in the controllable canonical form with unknown dynamics, i.e.,
\begin{equation}\label{eqn_dynamics}
	\begin{aligned}
		\dot{\bm{q}}_1=\bm{q}_2,\quad\dot{\bm{q}}_2=\bm{q}_3,\quad \cdots\quad \dot{\bm{q}}_m=\bm{f}(\bm{x})+\bm{u},
	\end{aligned}
\end{equation}
where $\bm{x} = [\bm{q}_1^T, \cdots, \bm{q}_m^T]^T \in \mathbb{X} \subset \mathbb{R}^{mn}$ denotes the state composed of vectors $\bm{q}_i \in \mathbb{R}^n$, $i = 1, \cdots, m$ with dimension $n\in\mathbb{N}_+$ and $\bm{u} \in \mathbb{R}^n$ is the control input.
This structure is found in many practical systems, e.g., in Euler-Lagrange dynamics such as robotic manipulators, drones, and underwater vehicles. 
While the structure of these systems is often known, accurate knowledge of the function $\bm{f}(\cdot)=[f_1(\cdot),\cdots,f_n(\cdot)]^T$, $f_i:\mathbb{X}\rightarrow\mathbb{R}$, $i=1,\cdots,n$ is usually not available, e.g., due to a high system complexity or unknown embedded environmental uncertainties.
Therefore, we assume $\bm{f}(\cdot)$ to be unknown, but pose the following assumption on it.

\begin{assumption}
	\label{assumption_f_bound}
	The unknown function $\bm{f}: \mathbb{X} \to \mathbb{R}^n$ is Lipschitz continuous on the compact domain $\mathbb{X}$ with Lipschitz constant $L_f$ induced by the Euclidean norm, i.e. $\|{\nabla} \bm{f}(\bm{x})\|\le L_f$.
\end{assumption}
This assumption or even stronger versions requiring global differentiability is commonly found since they ensure the existence of unique solutions of the differential equation \eqref{eqn_dynamics} \citep{khalil2015nonlinear}. 
The required local Lipschitz continuity is a very natural property of many physical systems which usually evolve smoothly over time. 
Therefore, \cref{assumption_f_bound} is not restrictive in practice.

\setlength{\abovedisplayskip}{5.0pt}
\setlength{\belowdisplayskip}{5.0pt}

The goal is to track a continuously differentiable and bounded reference trajectory $\bm{x}_d\!:\mathbb{R}_{0,+}\!\rightarrow\!\mathbb{X}$ with the state $\bm{x}$. 
In order to ensure the possibility of exact tracking, we assume reference trajectories of the form $\bm{x}_d\!=\![\bm{q}_{d,1}^T,\cdots,\bm{q}_{d,m}^T]^T$, $\bm{q}_{d,i}\!\in\!\mathbb{R}^n$, $i\!=\!1,\cdots,m$ defined by $\dot{\bm{q}}_{d,1} = \bm{q}_{d,2}$, $\cdots$, $\dot{\bm{q}}_{d,m} = \bm{q}_{d,m+1}$.
For achieving a high tracking accuracy, we employ a control law of the form 
\begin{align}\label{eqn_ctrl}
	\bm{u}(t)=\dot{\bm{q}}_{d,m}(t) - \hat{\bm{f}}(t) + \sum {_{i=1}^{m}} \bm{\Lambda}_{i} \bm{e}_i(t),
\end{align}
where $\bm{e}_i(t) = \bm{q}_{i}(t) - \bm{q}_{d,i}(t) \in \mathbb{R}^{n}, i = 1,2,\dots, m$ are tracking errors and $\bm{\Lambda}_i\in\mathbb{R}^{n\times n}$, $i=1,\cdots, m$ are control gain matrices.
The function $\hat{\bm{f}}(\cdot):\mathbb{R}_{0,+}\rightarrow\mathbb{R}^n$ is a compensation of the unknown function $\bm{f}(\bm{x}(\cdot))$.
This compensation can be realized by inferring a model $\bm{\mu}(\cdot):\mathbb{X}\rightarrow\mathbb{R}^n$ of $\bm{f}(\cdot)$, which requires the capability to generate training data as formalized in the following.

\begin{assumption}\label{assumption_mes_y}
	Noiseless measurements of the state $\bm{x}^{(\iota)}=\bm{x}(t_{\iota})$ and noisy measurements of the highest derivative $\bm{y}^{(\iota)} = \dot{\bm{q}}_m(t_{\iota}) + \bm{w}^{(\iota)}$ can be taken at arbitrary time instances $t_{\iota}$ with $\iota \in \mathbb{N}_0$. 
	The observation noise $\bm{w}^{(\iota)}\sim\mathcal{N}(\bm{0},\mathrm{diag}(\sigma_{o,1}^2, \cdots, \sigma_{o,n}^2))$ with $\sigma_{o,i} \in\mathbb{R}_+$, $i = 1,\cdots, n$ is assumed Gaussian, independent and identically distributed.
\end{assumption}
This assumption requires the exact measurement of the system state, which is a common requirement for the design of nonlinear control laws \citep{khalil2015nonlinear}. 
Moreover, it admits Gaussian perturbed observations of the derivative $\dot{\bm{q}}_m$, which for practical reasons often has to be approximated, e.g., using finite differences. 
Therefore, this assumption is commonly found when dealing with unknown dynamics \citep{koller2018learning,Greeff2021}. 
Note that relaxations to other distributions exist \citep{chowdhury2017kernelized}, but are out of the scope of this paper.
\looseness=-1

Moreover, \cref{assumption_mes_y} allows the generation of data during active control, such that a model $\bm{\mu}(\cdot)$ can be learned on-line. 
When real-world algorithms are used for this purpose, they need a considerable amount of time to update the model $\bm{\mu}_{\iota}(\cdot)$ using new data $(\bm{x}^{(\iota+1)},\bm{y}^{(\iota+1)})$\footnote{We use an index $\iota$ whenever necessary to emphasize the number of previously generated training samples.}. 
Moreover, the evaluation of such models at a state $\bm{x}$ often takes a non-negligible amount of time. 
Therefore, we cannot employ $\bm{\mu}(\bm{x}(t))$ as compensation $\hat{\bm{f}}(t)$ in practice, but need to work with the delayed value $\hat{\bm{f}}(t)=\bm{\mu}(\bm{x}(t_{\kappa(t)}))$, where
\begin{align}\label{eq:eval_times}
	t_{k+1}=t_k+\Delta(t_k),&& \kappa(t) = \argmax\limits {_{k\in\mathbb{N}}} ~ t_{k+1}<t
\end{align}
with the discretized time $t_k, \forall k \in \mathbb{N}, t_0 = 0$, and $\Delta:\mathbb{R}_{0,+}\rightarrow\mathbb{R}_+$ denotes the overall computation time which includes the time for determining a prediction 
in the off-line learning scenario in \cref{section_delayControl} and the time it takes to compute a model update and a prediction in the online learning problem in \cref{section_ET}. 
Note that the number of training samples $\iota$ at time $t$ does generally not correspond to the number of previous model evaluations $\kappa(t)$. 
In order to be able to determine the worst-case impact of the computation time $\Delta(\cdot)$ on the tracking accuracy, we require the following assumption.\looseness=-1

\begin{assumption} \label{assumption_delay}
	The overall computation time $\Delta(t)$ is bounded by a finite constant $\bar{\Delta} \in \mathbb{R}_+$ for all $t \in \mathbb{R}_{0,+}$ and its induced delay only affects the compensation $\hat{\bm{f}}(t)$.
\end{assumption}
Since a bounded computation time merely requires the termination of a learning algorithm after a finite number of computations, this assumption is satisfied by most practically employed methods.
Moreover, the limitation of the induced delay to the compensation $\hat{\bm{f}}(t)$ can be easily realized in practice by implementing the learned model in a parallel, non-blocking, process.
Therefore, \cref{assumption_delay} does not pose a severe restriction. 

Based on these assumptions, the controller \eqref{eqn_ctrl} yields the error dynamics
\begin{equation} \label{eqn_error_dynamics}
	\dot{\bm{e}}(t) = \bm{A} \bm{e}(t) + \bm{B} \big ( \bm{f}(\bm{x}(t)) - \bm{\mu}(\bm{x}(t_{\kappa(t)})) \big ),
\end{equation}
where $\bm{e}=\bm{x}-\bm{x}_d$, the matrices $\bm{A}\in \mathbb{R}^{mn \times mn}$ and $\bm{B}\in\mathbb{R}^{mn\times n}$ are defined as
\begin{align}\label{eqn_A}
	\bm{A} = \begin{bmatrix}
		\bm{0}_{(m-1)n \times n}	& \bm{I}_{(m-1)n}\\
		\bm{\Lambda}_1	& \left[\bm{\Lambda}_2, \cdots, \bm{\Lambda}_m \right]
	\end{bmatrix} &&
	\bm{B} = \begin{bmatrix}
		\bm{0}_{(m-1)n \times n}\\  \bm{I}_{n}
	\end{bmatrix}.
\end{align}

In this work, we consider the problem of ensuring an upper bound $\bar{e}\in\mathbb{R}_+$ for the tracking error $\bm{e}=\bm{x}-\bm{x}_d$ with high probability using a control law of the form \eqref{eqn_ctrl} by designing an event-triggered data selection strategy of the form
\begin{align} \label{eqn_data_selection}
	t_{\iota+1}=&\min {_{t_k> t_{\iota}}} t_k,
	&\text{such that } \|\bm{\mu}(\bm{x}(t_k))-\bm{f}(\bm{x}(t_k))\|\geq \upsilon^*(t_k,\bar{e},\bar{\Delta})
\end{align}
where $\upsilon^*(\cdot):\mathbb{R}\times\mathbb{R}_{+}\rightarrow\mathbb{R}_+$ denotes the trigger threshold function.

\section{Tracking Error Bounds with Delayed Gaussian Process Predictions} \label{section_delayControl}

Since the derivation of tracking error guarantees requires model error bounds, we consider Gaussian process regression as a machine learning method in this paper. 
The foundations of GP regression are outlined together with a prediction error bound in \cref{subsection_Preliminary_GP}. 
Based on this bound, tracking accuracy guarantees for control with delayed model predictions are derived in \cref{subsection_performance_no_trigger}.

\vspace{-0.1cm}
\subsection{Gaussian Process Regression} \label{subsection_Preliminary_GP}

Gaussian process regression is a modern machine learning technique, which can be used for the supervised inference of unknown functions.
For simplicity, we assume for the moment $n = 1$, such that the unknown function $f(\cdot)$ in \eqref{eqn_dynamics} is scalar. 
Then, a Gaussian process induces a distribution over $f(\cdot)$, denoted as $f(\cdot) \sim \mathcal{GP}(m(\cdot), k(\cdot,\cdot))$, which is completely specified by the prior mean $m:\mathbb{X}\rightarrow\mathbb{R}$ and the kernel $k:\mathbb{X} \times \mathbb{X} \rightarrow \mathbb{R}_{0,+}$. 
The prior mean $m(\cdot)$ can be used to incorporate parametric models into the regression.
Since they are often not available, a zero prior $m(\bm{x}) = 0$, $\forall\bm{x} \in \mathbb{X}$ is frequently used, which we also assume in the following without loss of generality. 
The kernel $k(\cdot,\cdot)$ reflects the prior covariance between evaluations of $f(\cdot)$ at different states $\bm{x}$ and can encode information such as periodicity or symmetry of $f(\cdot)$. 

When $N \in \mathbb{N}$ measurements $\bm{x}^{(\iota)}$, $y^{(\iota)}$, $\iota=1,\cdots, N$ satisfying \cref{assumption_mes_y} are available, this GP prior can be employed for regression using Bayesian principles. 
This merely requires the computation of the posterior distribution by conditioning the prior on the training data, which yields a Gaussian distribution with mean and variance
\begin{align} \label{eqn_GP_prediction}
	\mu(\bm{x}) = \bm{k}_{\bm{X}}^T(\bm{K}+\sigma_o^2\bm{I})^{-1}\bm{y}, && \sigma^2(\bm{x})=k(\bm{x},\bm{x})-\bm{k}_{\bm{X}}^T(\bm{K}+\sigma_o^2\bm{I})^{-1}\bm{k}_{\bm{X}},
\end{align}
at every test point $\bm{x}\in\mathbb{X}$.
For notational simplicity, the training targets $y^{(\iota)}$ are concatenated 
into a vector $\bm{y} = [y^{(1)}, \cdots, y^{(N)}]^T$ and the kernel gram matrix and vector are defined as $\bm{K} = [ k(\bm{x}^{(i)}, \bm{x}^{(j)}) ]_{i,j=1,\cdots,N}$ and $\bm{k}_X = [k(\bm{x}^{(1)}, \bm{x}), \cdots, k(\bm{x}^{(N)}, \bm{x})]^T$, respectively. 
The posterior mean $\mu(\cdot)$ can be used as a model of the unknown function $f(\cdot)$, while the variance $\sigma^2(\cdot)$ serves as a measure of epistemic uncertainty. 
In order to ensure that the epistemic uncertainty properly reflects the potential of model errors, we require the following assumption.\looseness=-1 
\begin{assumption}\label{assumption_sampleFun}
	The unknown function $f(\cdot)$ is a sample of a zero-mean prior GP with stationary, Lipschitz continuous and monotonically decreasing kernel $k(\cdot,\cdot)$.
\end{assumption}
A prior GP satisfying \cref{assumption_sampleFun} is flexible enough to represent the unknown function with suitable hyperparameters. 
Ensuring appropriate kernel structures can often be achieved in practice, e.g., by choosing universal kernels such as the squared exponential covariance function, which allows us to approximate continuous functions arbitrarily well \citep{Micchelli2006}.
The hyperparameters of such kernels can be obtained in practice via suitable hyperparameter tuning methods, e.g., \citep{Capone2021b}. 
Therefore, \cref{assumption_sampleFun} is not restrictive in practice. 

Based on this assumption, the following prediction error bound for GP regression can be shown.
\begin{lemma}[\cite{lederer2019uniform}] \label{lemma_uniform_error_bound}
	Consider an unknown function $f(\cdot)$ satisfying Assumptions~\ref{assumption_f_bound} and \ref{assumption_sampleFun}. 
	Moreover, assume that $N \!\in\! \mathbb{N}$ measurements $\bm{x}^{(\iota)}$, $y^{(\iota)}$ are available and satisfy \cref{assumption_mes_y}. 
	Then, for every $\delta \!\in\! \!(0,\!1)\! \!\subset\! \mathbb{R}$ and grid factor $\tau \!\in\! \mathbb{R}_+$, the prediction error of GP regression satisfies\looseness=-1
	\begin{align} \label{eqn_GP_error_bound}
		\mathbb{P}\left(| f(\bm{x}) - \mu(\bm{x}) | \le \eta_{\delta}(\bm{x}),~ \forall \bm{x}\in\mathbb{X}\right)\geq 1-\delta, && \eta_{\delta}(\bm{x}) = \sqrt{\beta_{\delta}} \sigma(\bm{x}) + \gamma_{\delta},
	\end{align}
	on a compact domain $\mathbb{X}\subset\mathbb{R}^{mn}$, where
	\begin{align}
		\beta_{\delta}=  2 \sum_{j=1}^{nm} \log{\left( \frac{\sqrt{nm}} {2 \tau} (\bar{x}_{j} - \underline{x}_{j}) + 1 \right)} - 2\log{\delta}, 
		&&\gamma_{\delta} = \left( \sqrt{\beta_{\delta}} L_{\sigma} + L_f + L_{\mu} \right) \tau,
	\end{align}
	with $\bar{x}_{j} = \max_{\bm{x} \in \mathbb{X}} x_j$, $\underline{x}_{j} = \min_{\bm{x} \in \mathbb{X}} x_j$, $x_j$ is the $j$-th dimension of $\bm{x}$, and $L_{\sigma}$ and $L_{\mu}$ denote the Lipschitz constants of standard deviation $\sigma(\cdot)$ and the mean $\mu(\cdot)$. 
\end{lemma}
This lemma allows us to bound the error of a learned GP model uniformly on a compact domain $\mathbb{X}$. 
The Lipschitz constants $L_{\sigma}$ and $L_{\mu}$ required for the error bound $\eta(\cdot)$ can be straightforwardly determined as shown in \citep{lederer2021gaussian}. 
Therefore, Lemma \ref{lemma_uniform_error_bound} provides a practically usable, state and data dependent error bound for models learned using GP regression.

\subsection{Performance Guarantees for Learning-Based Control with Delayed Predictions} \label{subsection_performance_no_trigger}

In order to apply GP regression for learning a model of the unknown function $\bm{f}(\cdot)\!=\![f_1(\cdot)\! \cdots\! f_n(\cdot)]^T$ in \eqref{eqn_dynamics} for the general case $n \geq 1$, we perform GP regression for each component $f_i(\cdot)$, $i=1,\cdots, n$, independently.
Afterwards, we concatenate the result in vectors $\bm{\mu}(\cdot)=[\mu_1(\cdot)\ \cdots\ \mu_n(\cdot)]^T$ and $\bm{\sigma}(\cdot)=[\sigma_1(\cdot)\ \cdots\ \sigma_n(\cdot)]^T$, where $\mu_i(\cdot)$ and $\sigma_i(\cdot)$ denote the mean and standard deviation of the GP trained using measurements of $f_i(\cdot)$. 
This allows us to directly employ the concatenated mean $\bm{\mu}(\cdot)$ in the control law \eqref{eqn_ctrl} by setting $\hat{\bm{f}}(t)=\bm{\mu}(\bm{x}(t_{\kappa(t)})$, where $t_{\kappa(t)}$ denotes the last model evaluation time defined through \eqref{eq:eval_times}. 
Since this allows us to bound the model error using Lemma \ref{lemma_uniform_error_bound}, we employ Lyapunov stability theory to analyze the tracking performance. 
For this purpose, we use a common quadratic  Lyapunov function $V(\bm{e}(t))=\bm{e}^T(t) \bm{P} \bm{e}(t)$ with a symmetric, positive definite matrix $\bm{P}\in\mathbb{R}^{mn\times mn}$, which is the solution of the continuous algebraic Riccati equation $\bm{A}^T \bm{P} + \bm{P} \bm{A} = - \bm{Q}$ for a given positive definite matrix $\bm{Q} \in \mathbb{R}^{m n \times m n}$.
This common Lyapunov function allows us to derive a tracking error bound as shown in the following theorem \footnote{Proofs for all theoretical results can be found at \url{https://mediatum.ub.tum.de/doc/1693032}.}. 

\begin{theorem} \label{theorem_UUB_no_trigger}
	Consider a system \eqref{eqn_dynamics}, where $\bm{f}(\cdot)$ is an unknown nonlinearity consisting of scalar functions $f_i(\cdot)$, $i=1,\cdots,n$, which satisfy Assumptions \ref{assumption_f_bound} and \ref{assumption_sampleFun}. 
	Assume that $n$ GPs are trained with $N\in\mathbb{N}$ measurements $\bm{x}^{(\iota)}$, $\bm{y}^{(\iota)}$ satisfying \cref{assumption_mes_y}, which results in the concatenated mean $\bm{\mu}(\cdot)$. 
	Assume that the computation time $\Delta(\cdot)$ of GP predictions satisfies \cref{assumption_delay}, such that we can employ the control law \eqref{eqn_ctrl} with $\hat{\bm{f}}(t)=\bm{\mu}(\bm{x}(t_{\kappa(t)})$ for $t_{\kappa(t)}$ defined in \eqref{eq:eval_times} and control gains $\bm{\Lambda}_i$ inducing a Hurwitz matrix $\bm{A}$ in \eqref{eqn_A}. 
	If the computation time bound $\bar{\Delta}$ satisfies $\bar{\Delta} < 1 / (2 L_f) $ and $\|\bm{e}(0)\| = 0$, then, the tracking error is bounded by
	\begin{align} \label{eqn_bound_no_update}
		\|\bm{x}(t)-\bm{x}_d(t)\|\leq \bar{e}=\chi \xi \big(2  L_f F \bar{\Delta} + \bar{\eta}_{\delta} \big), \quad \forall t\in\mathbb{R}_{0,+},
	\end{align}
	with probability of at least $2 (1 - \delta)^n - 1$, where $\xi = 2 \| \bm{P} \| \|\bm{Q}^{-1}\|$, $\chi^2 =\|\bm{P}^{-1}\| \|\bm{P}\|$, $\chi > 0$, $\bar{\eta}_{\delta} = \sup_{\bm{x} \in \mathbb{X}} \|\bm{\eta}_{\delta}\|$, $\bm{\eta}_{\delta}(\bm{x})=[\eta_{\delta,1}(\bm{x})\ \cdots\ \eta_{\delta,n}(\bm{x})]^T$ and
	\begin{align}
		F = \frac{1}{1 \!-\! 2 L_f \bar{\Delta}} &\Big(\! \| \bm{A} \| \sup_{\bm{x} \in \mathbb{X}} \| \bm{x} \|\! +\! \left\|\! \begin{bmatrix} \bm{\Lambda}_1& \cdots& \bm{\Lambda}_m\end{bmatrix}\!\right\| \sup_{t \in \mathbb{R}_{0,+}} \| \bm{x}_d(t) \| \!+ \!\sup_{t \in \mathbb{R}_{0,+}} \| \dot{\bm{q}}_{d,m}(t) \| \! + \! \bar{\eta}_{\delta} \!\Big).
	\end{align}
\end{theorem}

\cref{theorem_UUB_no_trigger} provides an intuitive insight into the sources of the tracking error. 
On the one hand, the accuracy of the GP model directly influences the tracking error as indicated by the model error bound $\bar{\eta}_{\delta}$, which can be reduced using more training data. 
On the other hand, lower computation time measured by $\bar{\Delta}$ can also decrease the tracking error bound $\bar{e}$. 
Since the computation time of GP predictions for $\bm{\mu}(\cdot)$ depends linearly ($\mathcal{O}(N)$) on the number of training samples $N$, these two values are inherently coupled, such that a suitable trade-off for the training set size $N$ must be found.

\section{Event-Triggered Online Learning under Computation Delays}\label{section_ET}

Since the tracking error bound \eqref{eqn_bound_no_update} increases with the computational delay and thus with 
the data set size in practice, data-efficiency is of crucial importance to ensure a high accuracy. Therefore, training samples should be carefully selected, which can be efficiently performed online by updating the model at the times when the model uncertainty is too high to guarantee the desired tracking error bound. This approach leads to the event-triggered online learning strategy derived in \cref{subsection_trigger_1}. Since such an online data generation process leads to potentially infinitely growing data sets, we analyze the effect of data deletion strategies on the derived guarantees in \cref{subsection_deletion}. In \cref{subsection_tradeoff}, we finally investigate the effect of the practically increased computation time due to online model updates to derive conditions when offline learning provides better tracking accuracy guarantees.

\subsection{Event-triggered Model Update Strategy} \label{subsection_trigger_1}

While the training data cannot directly be determined using \eqref{eqn_data_selection} due to the unknown dynamics $\bm{f}(\cdot)$, we can use the GP prediction error bound \eqref{eqn_GP_error_bound} as a proxy. 
This leads to the trigger condition
\begin{align} \label{eqn_trigger_condition_1}
	\|\bm{\eta}_{\delta}(\bm{x}(t_k))\| \geq \upsilon(t_k,\bar{e},\bar{\Delta}),
\end{align}
which can be interpreted as a conservative, but implementable version of the right side of \eqref{eqn_data_selection}.
Note that this trigger condition cannot be evaluated at arbitrary times $t\in\mathbb{R}_{0,+}$ as in previous approaches \citep{umlauft2019feedback, jiao2022backstepping} due to the considered non-negligible time required for the computation of the GP variance $\sigma^2(\bm{x})$ and thus $\|\bm{\eta}_{\delta}(\bm{x}(t_k))\|$. 
Instead, all computations are started at discrete times $t_k$, such that the trigger condition has only access to $\bm{x}(t_k)$, and the new evaluation of the GP mean $\bm{\mu}(\cdot)$ is available earliest at $t_k+\Delta(t_k)$. 
These restrictions need to be anticipated in the design of the event trigger as shown in the following theorem.\looseness=-1

\begin{theorem} \label{theorem_UUB_trigger_1}
	Consider a system \eqref{eqn_dynamics}, where $\bm{f}(\cdot)$ is an unknown nonlinearity consisting functions $f_i(\cdot), i \!=\! 1,\!\cdots\!,n$ which satisfies Assumptions \ref{assumption_f_bound} and \ref{assumption_sampleFun}.
	Assume that measurements $\bm{x}^{(\iota)}$, $\bm{y}^{(\iota)}$ satisfying \cref{assumption_mes_y} are used in each of the $n$ GPs, which results the concatenated mean $\bm{\mu}(\cdot)$ and standard deviation $\bm{\sigma}(\cdot)$.
	Assume that the computation time $\Delta(\cdot)$ required for a prediction and model update satisfies \cref{assumption_delay} and is bounded by $\bar{\Delta} \!<\! 1 / (2 L_f) $, such that the control law \eqref{eqn_ctrl} is employed with $\hat{\bm{f}}(t) \!=\! \bm{\mu}(\bm{x}(t_{\kappa(t)})$ for $t_{\kappa(t)}$ defined in \eqref{eq:eval_times} and control gains $\bm{\Lambda}_i$ inducing a Hurwitz matrix $\bm{A}$ in \eqref{eqn_A}.
	Pick $\bar{e} \!\ge\! 2 \chi \big( F \!+\! F_d \!+\! \xi L_f F \big) \bar{\Delta} \!+\! \chi \xi \underline{\eta}_{\delta}$ with $\underline{\eta}_{\delta} \!=\! \| [\underline{\eta}_{\delta,1}, \!\cdots\!, \underline{\eta}_{\delta,n}]^T \|$,  $\underline{\eta}_{\delta,i} \!=\! \sqrt{\beta_{\delta}} \sigma_{o,i} \!+\! \gamma_{\delta,i}$, and execute model updates at times $t_{\iota}$ defined by the event-trigger \eqref{eqn_trigger_condition_1} with threshold \looseness=-1
	\begin{align} \label{eqn_trigger_condition_u}
		\upsilon(t_k,\bar{e},\bar{\Delta}) = \xi^{-1} \max ( \| \bm{e}(t_k) \|, ~ \chi^{-1} \bar{e} ) - 2\big(\xi^{-1} (F + F_d) + L_f F \big)\bar{\Delta},
	\end{align}
	where $F_d \ge \| \dot{\bm{x}}_d(t) \|, \forall t \in \mathbb{R}_{0,+}$. Then, the tracking error is bounded by $\bar{e}$ with probability at least $2 (1 - \delta)^n - 1$, if the system initializes with $\| \bm{e}(0)\| \le \bar{e}$.
\end{theorem}
Due to the delayed impact of model updates, the trigger threshold \eqref{eqn_trigger_condition_u} cannot guarantee arbitrarily small tracking error bounds. Moreover, the time $\Delta(t_k)$ between two potential training samples prevents arbitrarily small prediction error bounds $\|\bm{\eta}_{\delta}(\bm{x})\|$, which also influences the achievable tracking error bound. Despite these restrictions, the trigger threshold \eqref{eqn_trigger_condition_u} exhibits the intuitive behavior that smaller desired error bounds yield more frequent model updates as indicated by a smaller value $\upsilon(t_k,\bar{e},\bar{\Delta})$. Moreover, a higher computation time bound $\bar{\Delta}$ has a similar effect as the trigger anticipates a potential future increase in the model and tracking error. Therefore, \cref{theorem_UUB_trigger_1} provides an effective method for ensuring a desired tracking error bound $\bar{e}$ in a practically relevant setting.\looseness=-1

\subsection{Tracking Error Guarantees with Data Deletion}\label{subsection_deletion}

When data is generated online using the event-trigger \eqref{eqn_trigger_condition_u}, the data set size $N$ can possibly grow infinitely, which would imply an unbounded computation time $\Delta(\cdot)$. 
Since this must be avoided in practice, a common strategy for online learning with GPs is the deletion of previous samples, such that the training set size $N$ and the computation time $\Delta(\cdot)$ remain bounded. 
This can be realized, e.g., by deleting the oldest data points \citep{Meier2016} or through information-theoretic criteria \citep{han2021stable}. 
Similarly as in \citep{umlauft2019feedback}, our guarantees for event-triggered online learning are not affected by such deletion strategies. 

\begin{corollary} \label{lemma_delete}
	Under the assumptions of \cref{theorem_UUB_trigger_1}, if model updates are executed at times $t_{\iota}$ defined through the event trigger \eqref{eqn_trigger_condition_1} with threshold \eqref{eqn_trigger_condition_u} after an arbitrary deletion strategy has been used for removing a data sample from the training set, then, the tracking error is bounded by $\bar{e}$ for all $\|\bm{e}(0)\|\le \bar{e}$ with probability at least $2(1-\delta)^n-1$. 
\end{corollary}
This corollary ensures that we can ensure \cref{assumption_delay} in practice by limiting the admissible data set size. 
While it allows the application of arbitrary deletion strategies without changes to the tracking error bound, the specific choice can still have a significant impact on the behavior of the control system. 
For example, poor deletion strategies, such as the deletion of all existing data, can significantly increase the number of triggered model updates. 
If the induced switching is excessive and the discrete steps in the control law become too large, real-world actuator can fail to follow the control signal with negative effects on the tracking accuracy. 
Therefore, a suitable method such as the established approach of deleting the oldest data point \citep{Meier2016} should be employed.
Note that the time for data deletion is directly considered in the overall computation time $\Delta(\cdot)$.
\looseness=-1

\subsection{Comparison between Offline Training and Online Model Updates} \label{subsection_tradeoff}

While we abstractly assume the existence of computation time bounds $\bar{\Delta}$ both for offline training in \cref{theorem_UUB_no_trigger} and for online updates in \cref{theorem_UUB_trigger_1}, the actual values of this bound significantly differ between these two scenarios in practice.
The reason for this is the requirement of the additional evaluation of the GP variance $\bm{\sigma}^2(\cdot)$ and the model update for event-triggered learning, which cause a quadratic complexity $\mathcal{O}(N^2)$. 
This leads to generally higher computation times $\Delta(\cdot)$ for online learning  in comparison to offline training, for which only mean evaluations ($\mathcal{O}(N)$) have to be computed online. 
Therefore, online learning cannot always be ensured to provide a benefit.

\begin{corollary} \label{theorem_tradeoff}
	Consider a system \eqref{eqn_dynamics}, where $\bm{f}(\cdot)$ is an unknown nonlinearity consisting of scalar functions $f_i(\cdot)$, $i=1,\cdots,n$, which satisfy Assumptions \ref{assumption_f_bound} and \ref{assumption_sampleFun}. Assume that the following two GP models can be employed in the control law \eqref{eqn_ctrl} with $\hat{\bm{f}}(t)=\bm{\mu}(\bm{x}(t_{\kappa(t)})$ for $t_{\kappa(t)}$ defined in \eqref{eq:eval_times} and control gains $\bm{\Lambda}_i$ inducing a Hurwitz matrix $\bm{A}$ in \eqref{eqn_A}:
	\begin{enumerate}
 \vspace{-0.14cm}
		\item an offline GP model trained with $N\in\mathbb{N}$ arbitrary measurements $\bm{x}^{(\iota)}$, $\bm{y}^{(\iota)}$ satisfying \cref{assumption_mes_y}, such that its computation time $\Delta(\cdot)$ for predictions is bounded by $\bar{\Delta}_1$ and its prediction error satisfies $\|\bm{\eta}_{\delta}(\bm{x})\| \leq \bar{\eta}_{\delta}, \forall \bm{x} \in \mathbb{X}$, which ensures a tracking error bound $\bar{e}_1$;
		\label{item:offlineGP}
  \vspace{-0.14cm}
		\item an online learning GP model updated with measurements $\bm{x}^{(\iota)}$, $\bm{y}^{(\iota)}$ satisfying \cref{assumption_mes_y} and defined through the event trigger \eqref{eqn_trigger_condition_1} with threshold \eqref{eqn_trigger_condition_u} and $\bar{e}_2 = 2 \chi \big( F + F_d + \xi L_f F \big) \bar{\Delta} + \chi \xi \underline{\eta}_{\delta}$ 
        in combination with an arbitrary data deletion strategy, such that the overall computation time $\Delta(\cdot)$ is bounded by $\bar{\Delta}_2$.
		\label{item:onlineGP}
	\end{enumerate}
 \vspace{-0.14cm}
	Assume that it holds that $\bar{\Delta}_1 \le \bar{\Delta}_2 < 1 / (2 L_f)$ and 
	\begin{align}\label{eqn_tradeoff}
		\bar{\Delta}_2 \ge \frac{\xi \tilde{\eta}_{\delta}}{2 (F + F_d)} 
		\quad \vee \quad
		\tilde{\Delta} \ge \frac{\xi \tilde{\eta}_{\delta} - 2 (F + F_d) \bar{\Delta}_1 }{2 (\xi L_f F + F + F_d)},
	\end{align}
	where $\tilde{\Delta}  = \bar{\Delta}_2 - \bar{\Delta}_1$ and $\tilde{\eta}_{\delta} = \bar{\eta}_{\delta} - \underline{\eta}_{\delta}$.
	Then, the guaranteed tracking error bound using the offline learning GP model \ref{item:offlineGP} is lower than 
	for the online learning GP model \ref{item:onlineGP}, i.e., $\bar{e}_1 \le \bar{e}_2$.
\end{corollary}
Since $\bar{e}_2$ in \cref{theorem_tradeoff} is the best guaranteeable tracking error bound using \cref{theorem_UUB_trigger_1}, this theorem implies that online learning does not offer a certifiable advantage over offline trained models if condition \eqref{eqn_tradeoff} holds. This condition crucially relies on the variable $\tilde{\eta}_{\delta}$, which measures the improvement in terms of the GP prediction error bound $\|\bm{\eta}_{\delta}\|$ due to event-triggered model updates. When the offline trained GP model is already highly accurate, event-triggered online learning can only provide a marginal improvement $\tilde{\eta}_{\delta}$. If the negative impact due to a computation time increase $\tilde{\Delta}$ exceeds this improvement as formalized on the right side of \eqref{eqn_tradeoff}, online learning can potentially yield worse tracking errors than offline trained models. Therefore, \cref{theorem_tradeoff} reflects the intuitive result that online learning is only meaningful if we do not have access to accurate offline models.
\section{Numerical Simulation} \label{section_simulation}

To illustrate the effect of the computation delay on the control performance, we evaluate the 
proposed event-triggered model update strategy in a numerical simulation using the system
dynamics $\dot{x}_1 = x_2$, $\dot{x}_2 = f(\bm{x})  + u = \sin(x_1) + 0.5\big(1 + \exp (x_2 / 10) \big)^{-1} + u$, where $\bm{x} = [x_1, x_2]^T \in \mathbb{X} = [-1.5,1.5] \times [-1.5,1.5] \subset \mathbb{R}^2$.
The desired trajectory is set to $x_{d,1}(t) = \sin(t), x_{d,2}(t) = \cos(t)$ satisfying \cref{assumption_delay}. 
The control gains $\Lambda_i$ in \eqref{eqn_ctrl} is chosen as $\Lambda_1 = \Lambda_2 = -2$, and the symmetric positive definite matrix is selected as $\bm{Q} = \bm{I}_2$. 
For learning, we use Gaussian processes with squared exponential kernel $k(\bm{x}, \bm{x}') = \sigma_f^2 \exp (-0.5 l^{-2} (\bm{x} - \bm{x}')^T (\bm{x} - \bm{x}')  )$, where $\sigma_f = 1$ and $l = 0.2$. 
All training samples $y^{(i)}$ are perturbed by Gaussian noise with standard deviation $\sigma_o = 0.01$. 
The simulation time is set to $20$ and every simulation is repeated $100$ times to account for the randomness in training data. 
Unless stated otherwise, all simulations are started with an initial data set containing $N_0 = 100$ samples evenly distributed in $\mathbb{X}$.

Based on these data sets, we first illustrate the effect of computation delays on the tracking error and investigate the necessary trade-off between model accuracy and computation time in  \cref{subsection_delay_influence_illustration}. 
In \cref{subsection_tradeoff_illustration}, the  proposed event-triggered model update strategy is demonstrated.

\subsection{Influence of Delay on Tracking Errors} \label{subsection_delay_influence_illustration}

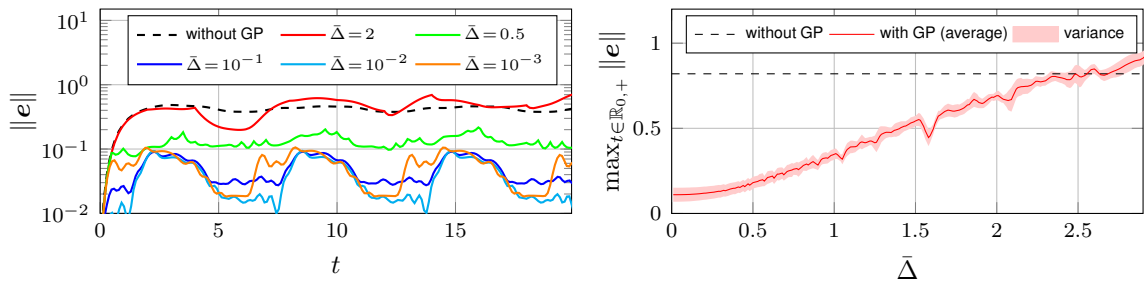
\begin{figure}[t] 
	\centering
	\begin{tikzpicture}
		\def\file{figures/Delay.txt}
		\begin{semilogyaxis}[xlabel={$t$},ylabel={$\|\bm{e}\|$},
			xmin=0, ymin = 0.01, xmax = 19.9,ymax=15,legend columns=3,
			width=0.515\textwidth,height=4.3cm,legend pos= north east,
			ylabel shift = -0.2cm,  xshift=-7.6cm]
			\addplot[black, thick, dashed]    table[x = t0 , y  = e0 ]{\file};
			\addplot[red, thick]    table[x = t1 , y  = e1 ]{\file};
			\addplot[green, thick]    table[x = t2 , y  = e2 ]{\file};
			\addplot[blue, thick]    table[x = t3 , y  = e3 ]{\file};
			\addplot[cyan, thick]    table[x = t4 , y  = e4 ]{\file};
			\addplot[orange, thick]    table[x = t5 , y  = e5 ]{\file};
			\legend{$\!\!$ without GP $~$, $\!\!$ $\bar{\Delta} \!=\! 2 $ $~$, $\!\!$ $\bar{\Delta} \!=\! 0.5 $ $~$, $\!\!$ $\bar{\Delta} \!=\! 10^{-1} $ $~$, $\!\!$ $\bar{\Delta} \!=\! 10^{-2} $, $\!\!$ $\bar{\Delta} \!=\! 10^{-3} $}
		\end{semilogyaxis}
		\def\file{figures/DelayMontCarlo.txt}    
		\begin{axis}[xlabel={$\bar{\Delta}$},ylabel={$\max_{t\in\mathbb{R}_{0,+}} \|\bm{e}\|$},
			xmin=0, ymin = 0, xmax = 2.9, ymax = 1.2,legend columns=3,
			width=0.515\textwidth,height=4.3cm,legend pos= north west,
			ylabel shift = -0.2cm,  xshift=0cm]
			\addplot[black, dashed]    table[x = bx , y  = by ]{\file};
			\addplot[red]    table[x = t , y  = e ]{\file};
			\addplot+[name path=max,black,no markers, draw=none] table[x = t , y  = max_e ]{\file};
			\addplot+[name path=min,black,no markers, draw=none] table[x = t , y  = min_e ]{\file};
			\addplot[red!20] fill between[of=max and min];
			\legend{{without GP},{with GP (average)},,, {variance}}
		\end{axis}
	\end{tikzpicture}
	\vspace{-1cm}
	\caption{Tracking errors without online learning under different computational delays (left).	Dependency of the maximum tracking error on the computational delay (right). 
	}
	\label{figure_delay_illustration}
\end{figure}

In order to investigate the effect of the computation delay $\Delta(t)$ on the control performance, we first consider GP models for fixed data sets, i.e., with only $N_0$ initial samples and without model updates.
For simplicity, we assume constant computation times $\Delta(t)\!=\!\bar{\Delta}, \forall t \!\in\! \mathbb{R}_{0,+}$ which are independent of $N_0$, such that the effect of available computational resources can be investigated.
The resulting tracking errors for different delays $\bar{\Delta}$ are depicted on the left of \cref{figure_delay_illustration}. For large values such as $\bar{\Delta}\!=\!2$   
\begin{wrapfigure}[11]{r}{0.5\linewidth}
	\vspace{-0.2cm}
	\centering
	\begin{tikzpicture}
		\def\file{figures/TradeOff.txt}
		\begin{axis}[xlabel={$N_0$},ylabel={$\max_{t\in\mathbb{R}_{0,+}} \| \bm{e} \|$},
			xmin=0, ymin = 0.1, xmax = 199.9,ymax=0.65,legend columns=1,
			width=7.5cm,height=4.3cm,legend pos=north east]
			\addplot[red]    table[x = N , y  = e ]{\file};
			\addplot+[name path=max,black,no markers, draw=none] table[x = N , y  = max_e ]{\file};
			\addplot+[name path=min,black,no markers, draw=none] table[x = N , y  = min_e ]{\file};
			\addplot[red!20] fill between[of=max and min];
			\draw [stealth-, very thick] (90,0.1765) -- (90,0.32) node[pos = 1, above, font = \small]{Minimum};
			\legend{{average},,,{empirical variance}}
		\end{axis}
	\end{tikzpicture}
	\vspace{-0.6cm}
	\caption{
		Accuracy-delay trade-off due to a joint dependence on the training set size $N_0$.\looseness=-1
	}
	\label{figure_Tradeoff}
\end{wrapfigure}
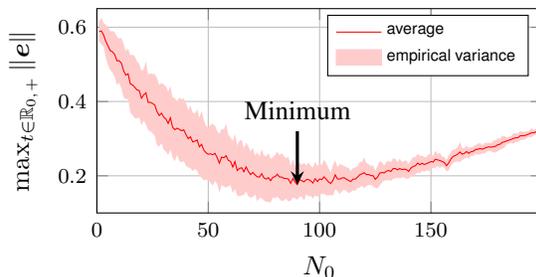 
and $\bar{\Delta}\!=\!0.5$,
the performance is strongly affected by the computation delay and can even exceed that of a controller without any GP model, i.e., $\hat{\bm{f}}(t) \!=\! \bm{0}, \forall t \!\in\! \mathbb{R}_{0,+}$. In contrast, a 
decrease of the delay below $\bar{\Delta}\!=\!10^{-1}$ has only a marginal effect on the tracking error. This effect is in line with \cref{theorem_UUB_no_trigger}, which guarantees a linear dependency of the tracking error bound 
$\bar{e}$ on both the prediction error bound $\bar{\eta}_{\delta}$ and the delay $\bar{\Delta}$. When $\bar{\eta}_{\delta}$ is kept constant, this implies the observed vanishing impact of a delay reduction. 
As illustrated on the right side of \cref{figure_delay_illustration}, we can even see the linear growth of the maximum error for sufficiently large delays, which is guaranteed for the error bound $\bar{e}$ in \eqref{eqn_bound_no_update}.\looseness=-1

In addition to this linear dependency of the error on the delay bound $\bar{\Delta}$, we can also directly observe the accuracy-delay trade-off discussed in \cref{section_delayControl}. For this purpose, we employ the knowledge of the linear GP prediction complexity to model the computational delay as $\bar{\Delta}=c N_0$, where we exemplarily choose $c = 0.05$ and $N_0$ in the range between 0 and 200.
The resulting dependency of the maximum tracking error on the number of training samples is depicted in \cref{figure_Tradeoff}. It can be clearly seen that the error reduces at first due to the improved accuracy.
After $N_0 \approx 90$ data points have been reached, the negative effect of the delay becomes dominant, such that the error starts to increase. This leads to a clearly visible minimum, at which the GP accuracy and the caused computational delay are optimally balanced. 

\subsection{Control Performance with Event-triggered Update} \label{subsection_tradeoff_illustration}

\begin{figure}[t] 
	\centering
	\begin{tikzpicture}
		\def\file{figures/Update.txt}
		\begin{semilogyaxis}[xlabel={$t$},ylabel={$\|\bm{e}\|$},
			xmin=0, ymin = 0.0001, xmax = 19.9,ymax=20,legend columns=2,
			width=0.515\textwidth,height=4.3cm,legend pos= north east,
			ylabel shift = -0.2cm,, xshift=-7.6cm]
			\addplot[black, thick,line width=1, dashed]    table[x = t0 , y  = e0 ]{\file};
			\addplot[red, thick]    table[x = t1 , y  = e1 ]{\file};
			\addplot[green, thick]    table[x = t2 , y  = e2 ]{\file};
			\addplot[blue, thick]    table[x = t3 , y  = e3 ]{\file};
			\legend{{$\!\!$ $\bar{\Delta} \!=\! 0.01$, without update $~$}, $\!\!$ $\bar{\Delta} \!=\! 0.01$ $~$, $\!\!$ $\bar{\Delta} \!=\! 0.1$ $~$, $\!\!$ $\bar{\Delta} \!=\! 0.45$ $~$, $\!\!$ $\bar{\Delta} \!=\! 0.25$ $~$, $\!\!$ $\bar{\Delta} \!=\! 0.5$}
		\end{semilogyaxis}
		\def\file{figures/ThresholdMontCarlo.txt}    
		\begin{axis}[xlabel={$\tilde{\Delta}$},ylabel={$\max_{t\in\mathbb{R}_{0,+}} \| \bm{e} \|$},
			xmin=0.01, ymin = 0, xmax = 0.48,ymax=0.22,legend columns=2,
			width=0.515\textwidth,height=4.3cm,legend pos= north west,
			yticklabels={,$0$,$0.05$,$0.10$,$0.15$},
			ylabel shift = -0.2cm, xshift=0cm]
			\addplot[black, dashed]    table[x = tb , y  = b ]{\file};
			\addplot+[name path=bmax,black,no markers, draw=none] table[x = tb , y  = bmax ]{\file};
			\addplot+[name path=bmin,black,no markers, draw=none] table[x = tb , y  = bmin ]{\file};
			\addplot[black!20] fill between[of=bmax and bmin];
			\addplot[red]    table[x = t , y  = e ]{\file};
			\addplot+[name path=emax,black,no markers, draw=none] table[x = t , y  = emax ]{\file};
			\addplot+[name path=emin,black,no markers, draw=none] table[x = t , y  = emin ]{\file};
			\addplot[red!20] fill between[of=emax and emin];
			\legend{{$\bar{\Delta}=0.01$, without update (average)},,,{variance}, {with update (average)},,,{variance}}
		\end{axis}
	\end{tikzpicture}
	\vspace{-1cm}
	\caption{
		(Left) Tracking errors with event-triggered online learning under different computational delays.
        Overall, 595 of 2077, 197 of 900 and 115 of 118 new samples are online selected for model update with $\bar{\Delta} \!=\! 0.01$, $0.1$ and $0.45$, respectively.
        (Right) Dependency of the maximum tracking error on the additional delay $\tilde{\Delta}$ caused by model updates in comparison to offline trained GPs with fixed computation time.\looseness=-1
	}
	\label{figure_update_illustration}
\end{figure}
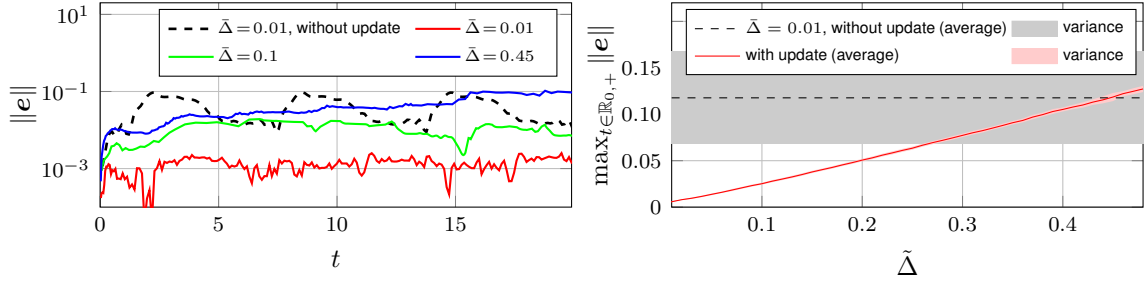

We demonstrate the effectiveness of the proposed online learning strategy by choosing the minimal, admissible error bound $\bar{e} \!=\! 2 \chi \big( F \!+\! F_d \!+\! \xi L_f F \big) \bar{\Delta} \!+\! \chi \xi \underline{\eta}_{\delta}$ in the trigger threshold \eqref{eqn_trigger_condition_u}. 
Due to the quadratic complexity of GP updates, we model the computation time as $\Delta(t)\!=\!c N^2(t)$, where $N(t)$ denotes the number of training samples at time $t$. In order to ensure a bounded computation time, we delete the oldest data point once the data set reaches the threshold $\bar{N}\!=\!200$, yielding $\bar{\Delta}\!=\!c\bar{N}^2$. Examples of the tracking errors resulting from different values of $c$ are depicted on the left side of \cref{figure_update_illustration}. It can be clearly seen that larger delay bounds $\bar{\Delta}$ result in higher errors. When comparing to an offline trained GP model with computational delay $\bar{\Delta}\!=\!0.01$, we can even observe that the growth in computation time due to online learning can increase the error, which is guaranteed to happen for the tracking error bound due to \cref{theorem_tradeoff}. In fact, this effect consistently occurs with increasing difference $\tilde{\Delta}$ between computation times for the considered random initial data sets, as shown on the right side of \cref{figure_update_illustration}. This is due to the linear growth of the error with event-triggered learning, which corresponds to the behavior of the tracking error bound guaranteed by \cref{theorem_UUB_trigger_1}. Therefore, all effects derived for the tracking error bounds can be observed in the actual tracking errors, which demonstrates the practical importance of the derived theoretical guarantees.

\section{Conclusions} \label{section_conclusion}

In this paper, the effect of computational delay caused by GP models on the control performance is analyzed, which reveals a trade-off between prediction accuracy and computational delay.
An event-triggered online model update strategy under computational delays is proposed, and we show that it can offer advantages over offline trained GP models.
Finally, simulations demonstrate the effectiveness of the derived theories and the event-triggered online learning strategy.

\acks{
	This work is supported by the Federal Ministry of Education and Research of Germany in the programme of “Souverän. Digital. Vernetzt.” under the joint project 6G-life with identification number: 16KISK002, and by the European Research Council (ERC)  Consolidator  Grant  ”Safe  data-driven  control  for human-centric systems (CO-MAN)” under grant agreement number 864686, and by the European Union’s Horizon 2020 research and innovation programme under grant agreement number 871295.
}

\bibliography{yourbibfile}

\end{document}